\documentclass[3p]{elsarticle}

\usepackage{hyperref}
\usepackage{tabularx}
\usepackage{bigstrut}
\usepackage{booktabs}
\usepackage{bm}
\usepackage{physics}
\usepackage{comment}
\usepackage{mathtools}
\usepackage{color}

\journal{Annals of Physics}

\begin{document}

\begin{frontmatter}

\title{Beyond the universal Dyson singularity for 1-D chains with hopping disorder}

%\author{Akshay Krishna, R. N. Bhatt}
%\address{Department of Electrical Engineering, Princeton University, Princeton NJ 08544, USA}

\author[mymainaddress,mysecondaryaddress]{Akshay Krishna}
%\ead{akshaykri.com}

\author[mymainaddress]{R. N. Bhatt}
%\ead{ravin@princeton.edu}

\address[mymainaddress]{Department of Electrical Engineering, Princeton University, Princeton NJ 08544, USA}
\address[mysecondaryaddress]{KLA Corporation, 2350 Green Road Suite 100, Ann Arbor, MI, 48105, USA}

\begin{abstract}
We study a simple non-interacting nearest neighbor tight-binding model in one dimension with disorder, where the hopping terms are chosen randomly.
This model exhibits a well-known singularity at the band center both in the density of states and localization length. 
If the probability distribution of the hopping terms is well-behaved, then the singularities exhibit universal behavior, the functional form of which was first discovered by Freeman Dyson in the context of a chain of classical harmonic oscillators.
We show here that this universal form can be violated in a tunable manner if the hopping elements are chosen from a divergent probability distribution.
We also demonstrate a connection between a breakdown of universality in this quantum problem and an analogous scenario in the classical domain -- that of random walks and diffusion with anomalous exponents.
\end{abstract}

\end{frontmatter}

%\linenumbers

\section{Introduction}

The question ``How do particles move?'' is a central one in physics, and one that has played a key role in the development of modern physics over the centuries.
The work of Galileo and Newton on ballistic motion in the classical domain was perhaps the progenitor of the study of \emph{transport}.
In the early twentieth century, the study of transport broadened in two orthogonal directions.
The first was the extension of transport from perfect translationally invariant systems to random systems.
This was accompanied by developments in the theory of diffusion, random walks and Brownian motion.
Second, the discovery of quantum mechanics led to a revolution in the theory of transport at the very smallest scales.
In time, quantum mechanical and semi-classical transport came to be the backbone of a new area of solid state physics, distinct from the cornerstone provided by Bloch's theorem.

The combination of simultaneous randomness and quantumness was first studied in 1958 by Philip Anderson in a tight-binding model of spinless electrons in a three-dimensional lattice  \citep{Anderson1958}, where he found the surprising fact that electrons could localize in such conditions.
In time, it was realized that disorder generically induces a quantum phase transition from a metallic phase to an insulating one  \citep{Mott1961} .
This metal-insulator transition is accompanied by an alteration in the nature of the eigenstates of the Hamiltonian -- from extended Bloch-like states in the metallic phase to localized states with exponentially decaying envelopes in the insulating phase.
As the disorder is ramped up and the transition is crossed, conductance drops and the transmission coefficients go to zero.

Our current understanding of Anderson transitions is built on the scaling theory of localization  \citep{GangofFour1979, Lee1985, Evers2008}, and its applicability to the canonical universality classes in various dimensions.
These universality classes are based on the existence of three discrete symmetries, namely, time-reversal, particle-hole and sublattice (chiral) symmetry  \citep{Chiu2016}.
The Hamiltonian of the standard Anderson model consists of two kinds of terms -- a random uncorrelated on-site potential, and a constant nearest neighbor hopping term from discretizing the kinetic energy on a lattice.
This Hamiltonian lies in the orthogonal symmetry class, as it obeys time-reversal and $U(1)$ spin-rotation symmetry. 
For this symmetry class, the lower critical dimension is two.
Hence, in $d=1$ and $d=2$ dimensions, infinitesimally small disorder causes Anderson localization of the entire spectrum.

However, in the last few decades, many tight-binding models in one dimension have been developed that do not obey this standard picture. 
These models often involve significant modifications to the Anderson Hamiltonian that push it into a different symmetry class and lead to more complicated behavior.
Examples of such modifications include: introducing correlations in the disorder  \citep{Phillips1990, Moura1998, Izrailev1999, Moura1999}, adding long-range hopping  \citep{Zhou2003}, and truncating the Hilbert space  \citep{Krishna2018}.

In this paper, we discuss another such modification: the model with pure nearest-neighbor hopping disorder with no on-site potential.
The only non-zero matrix elements of the Hamiltonian are along the first diagonals above and below the principal diagonal, and hence it is also known as the off-diagonal disorder model \citep{Theodorou1976, Eggarter1978}.

One consequence of having no (or a constant) on-site potential is that the Hamiltonian becomes bipartite -- every odd numbered site is only coupled to even-numbered sites, and vice versa.
Technically, this is a form of sublattice (chiral) symmetry, and leads to an exact pairing of states in the spectrum.
Every eigenstate $\ket{\psi_+}$ with energy $E$ is accompanied by another eigen state $\ket{\psi_-}$ with energy $-E$, such that $\ket{\psi_+} + \ket{\psi_-}$ has support only on even-numbered sites, and $\ket{\psi_+} - \ket{\psi_-}$ has support only on odd sites.
For this reason, in the rest of the paper, we assume $E \geq 0$ without any loss of generality.

\section{The disordered hopping model}

The problem we study traces its genesis to Freeman Dyson's work on a classical one-dimensional chain of random harmonic oscillators with Poisson distributed couplings \citep{Dyson1953}.
He found a peculiar singularity in the distribution function of the frequencies of normal modes of vibration of this model.
In 1970, Smith mapped that the problem onto the nearest neighbor spin-1/2 XY chain with random couplings whose squares have a generalized Poisson distribution.
He showed that the low temperature thermodynamic behavior of this system has a very similar singularity \citep{Smith1970}.
Applying the Jordan-Wigner transformation  \citep{Lieb1961} converts the XY model to a spinless fermionic chain with nearest neighbor coupling.
This model was shown to have a singular density of states at the center of the band  \citep{Theodorou1976}.
Soon after, Eggarter and Riedinger found that any well-behaved distribution\footnote{We use the term `well-behaved' somewhat loosely to encompass commonly occurring distributions in physics such as Gaussian, Poisson, and box distributions, and more generally distributions whose all moments are finite. Past studies of this model in the literature have implicitly confined themselves to such cases only.} of hoppings  \citep{Eggarter1978, Dhar1980} led to a density of states divergence of the form $\rho (E) \sim 1/|E \ln^3 |E||$.
This divergence in the density of states is accompanied by a divergence in the localization length, $\xi(E) \sim |\ln |E||$.
The state at zero energy is not conventionally extended, however, and decays with an envelope $\psi(r) \sim \exp \left( -\sqrt{r/r_0} \right)$  \citep{Fleishman1977, Soukoulis1981}.
This band center anomaly is well documented in the literature  \citep{Stone1981, Ziman1982, Evangelou1986, Roman1987, Markos1988, Roman1988, Inui1994}.
In spite of its simplicity and long history, this model remains an active topic of research, with many new results in the past decade \cite{Mard2014, Zhao2015, Bera2016, Stamatiou2020}.
In honor of Freeman Dyson's pioneering work of 1953, we refer to this universal behavior as the `Dyson' singularity.
%For brevity, we refer to the class of one-dimensional disordered nearest-neighbor hopping Hamiltonians with non-singular hopping distributions\footnote{Distributions of hopping $t$ that are less singular than the universal form $p(t) \sim 1/(t \ln^2(1/t))$ are also included} and the resultant singularity in $\rho (E)$ and $\xi(E)$ as the `Dyson' class and singularity respectively, in light of Freeman Dyson's pioneering work on the topic.

Like in previous studies, our Hamiltonian is that of spinless fermions hopping on a one-dimensional lattice, with nearest neighbor hopping \begin{align}
    H &= \sum_i t_i (c^\dagger_i c_{i+1} + c^\dagger_{i+1} c_{i}).
\end{align}
The fermionic creation and annihilation operator on site $i$ are denoted by $c^\dagger_i$ and $c_i$ respectively, and the hopping terms $t_i$ are non-negative.
The key feature of our study is that the $t_i$ are independent identically-distributed random variables drawn from a probability distribution that is \emph{not} well-behaved.
The `not-well-behavedness' of the hoppings is made more concrete by specifying that their probability distribution to be of the form \begin{align}
    p(t) \simeq \begin{cases} \frac{b_{0}}{t {|\ln t|}^{{\alpha_0}+1}}, \qquad t \to 0^+ \\
    \frac{b_{\infty}}{t {|\ln t|}^{{\alpha_\infty}+1}}, \qquad t \to +\infty.
    \end{cases} \label{eq:p_alpha}
\end{align}
The index $\alpha_0 > 0$ controls the strength of the divergence at the origin, and $\alpha_\infty > 0$ controls the thickness of the tail of the distribution at large values.
The coefficients $b_{0}$ and $b_{\infty}$ are non-negative scale factors that cannot both be zero simultaneously.
In our recent paper \citep{Krishna2020}, we only considered the case where $b_{\alpha_\infty} = 0$, so that the `not-well-behavedness' of the distribution came solely from the form of its divergence at the origin.
In that case, all moments of the hopping $t$ are finite (if the distribution does not have fat power-law tails at infinity).
However, that requirement can be relaxed to extend the non-universal results described in  \citep{Krishna2020} to a broader class of distributions, as we discuss here.

The logarithm of the hoppings ($u \equiv \ln t$) is a frequently occurring quantity in the ensuing discussion.
Its distribution is found by a simple change of variable from Eq.\ \eqref{eq:p_alpha}, and is given by simple power laws \begin{align}
    p(u) \simeq \begin{cases} \frac{b_{0}}{{|u|}^{{\alpha_0}+1}}, \qquad u \to -\infty \\
    \frac{b_{\infty}}{{u}^{{\alpha_\infty}+1}}, \qquad u \to +\infty.
    \end{cases} \label{eq:p_u}
\end{align}
These power-law tails lead to divergences in the moments of $u$, and are the root cause of non-universal behavior we see in this model.
In particular, let the variance of $u$ be denoted $\sigma_u^2$.
This may  be infinite if $\alpha_0 < 2$ or $\alpha_\infty < 2$.

While the case of infinitesimally small hopping ($t \to 0$ or $u \to -\infty $) is a physically realistic condition that may occur when nearest neighbors are very weakly coupled (for instance if the inter-nuclear spacing between atoms $r \to \infty$), the opposite limit ($t \to \infty$) is probably more mathematical in nature.
Large hoppings can be realized when the effective mass is low, or the inter-nuclear separation very small, however, a divergent hopping would be precluded by the existence of a length-scale cut-off.

\section{The zero energy state} \label{sec:zero}

The Schrodinger equation for the Hamiltonian \eqref{eq:p_alpha} may be written in terms of the on-site wave function amplitudes $\psi_i$ as \begin{align}
    t_i \psi_{i+1} + t_{i-1}\psi_{i-1} = E \psi_i. \label{eq:sch}
\end{align}
At $E=0$, this allows us to solve the recurrence relation above very simply for the wave function amplitude \begin{align}
    \psi_{2n} &= (-1)^n \left( \frac{{\prod\limits_{i \text{ odd}}^{2n} t_i} }{ {\prod\limits_{i \text{ even}}^{2n} t_i}} \right) \psi_0.
\end{align}
The rate of growth of the wave function's envelope can be determined by \begin{align}
      \ln \left| \frac{\psi_{2n}}{\psi_0} \right| = \sum_{i \text{ odd}}^{2n} \ln t_i - \sum_{i \text{ even}}^{2n} \ln t_i = \sum_{i \text{ odd}}^{2n} u_i - \sum_{i \text{ even}}^{2n} u_i. \label{eq:env}
\end{align}

The statistical properties of the wave function envelope in the long-chain limit ($n \to \infty$) can be determined by appealing to the central limit theorem (CLT) and its generalized version  \citep{Gnedenko1968, Shintani2018}.
The CLT specifies that the form of the limiting distribution for sums of random variables is a Gaussian $\mathcal{N}(\mu, \sigma^2)$, with some mean $\mu$ and standard deviation $\sigma$, so long as each of the random variables has finite variance.
When the variance is infinite, as is the case when $\alpha_0, \alpha_\infty < 2$, then the generalized CLT provides the functional form of the limiting distributions in terms of the \emph{L\'{e}vy alpha-stable distributions}.
These distributions are a family of normalized probability density functions that do not generally have closed form expressions \cite{Rocha2019}, except for a few special cases.
The L\'{e}vy alpha-stable distributions are usually denoted by $\mathcal{S}(\alpha, \beta, \gamma, \delta)$, where the four parameters, namely, $\alpha \in (0, 2]$, $\beta \in [-1, 1]$, $\gamma \in (0, \infty)$ and $\delta \in (-\infty, \infty)$ specify the stability, skewness, scale and location respectively. 
Applying the CLT or generalized CLT to Eq.\ \eqref{eq:env}, it follows that \begin{align}
   &\lim\limits_{n \to \infty} \frac{1}{\sqrt{2n}} \ln \left| \frac{\psi_{2n}}{\psi_0} \right| \quad \xrightarrow[]{\text{dist.}} \quad \mathcal{N}(0, \sigma_u^2), \qquad \qquad \quad  \text{if } \alpha_0 > 2 \text{ and } \alpha_\infty > 2. \label{eq:CLT1} \\
   &\lim\limits_{n \to \alpha} \frac{1}{(2n)^{1/\alpha}} \ln \left| \frac{\psi_{2n}}{\psi_0} \right| \quad \xrightarrow[]{\text{dist.}} \quad \mathcal{S} \left(\alpha, 0, \gamma(\alpha), 0 \right), \qquad \text{otherwise.} \label{eq:CLT2}
\end{align}

The notation $\xrightarrow[]{\text{dist.}}$ denotes that, mathematically speaking, this is a convergence in distribution, not pointwise convergence  \citep{Gnedenko1968}.
In the case where one or both of $\alpha_0$ and $\alpha_\infty$ are smaller than 2, the variance of $u$, denoted by $\sigma_u^2$, is infinite, and the properties of system are controlled by the smaller of the two.
We denote this parameter as $\alpha \equiv \min (\alpha_0, \alpha_\infty)$.
The scale parameter $\gamma(\alpha)$ is \begin{align}
    \gamma(\alpha) = \left(\frac{\pi b}{2\alpha \sin(\frac{\pi \alpha}{2}) \Gamma(\alpha)} \right)^{\frac{1}{\alpha}}, \qquad \text{where } \quad  b \equiv \begin{cases}
    b_0, &\text{if } a_0 < a_\infty, \\
    b_\infty, &\text{if } a_0 > a_\infty, \\
    b_0 + b_\infty, &\text{if } a_0 = a_\infty.
    \end{cases} \label{eq:gamma}
\end{align}

\begin{figure}[t!]
	\centering
		\includegraphics[scale=.7]{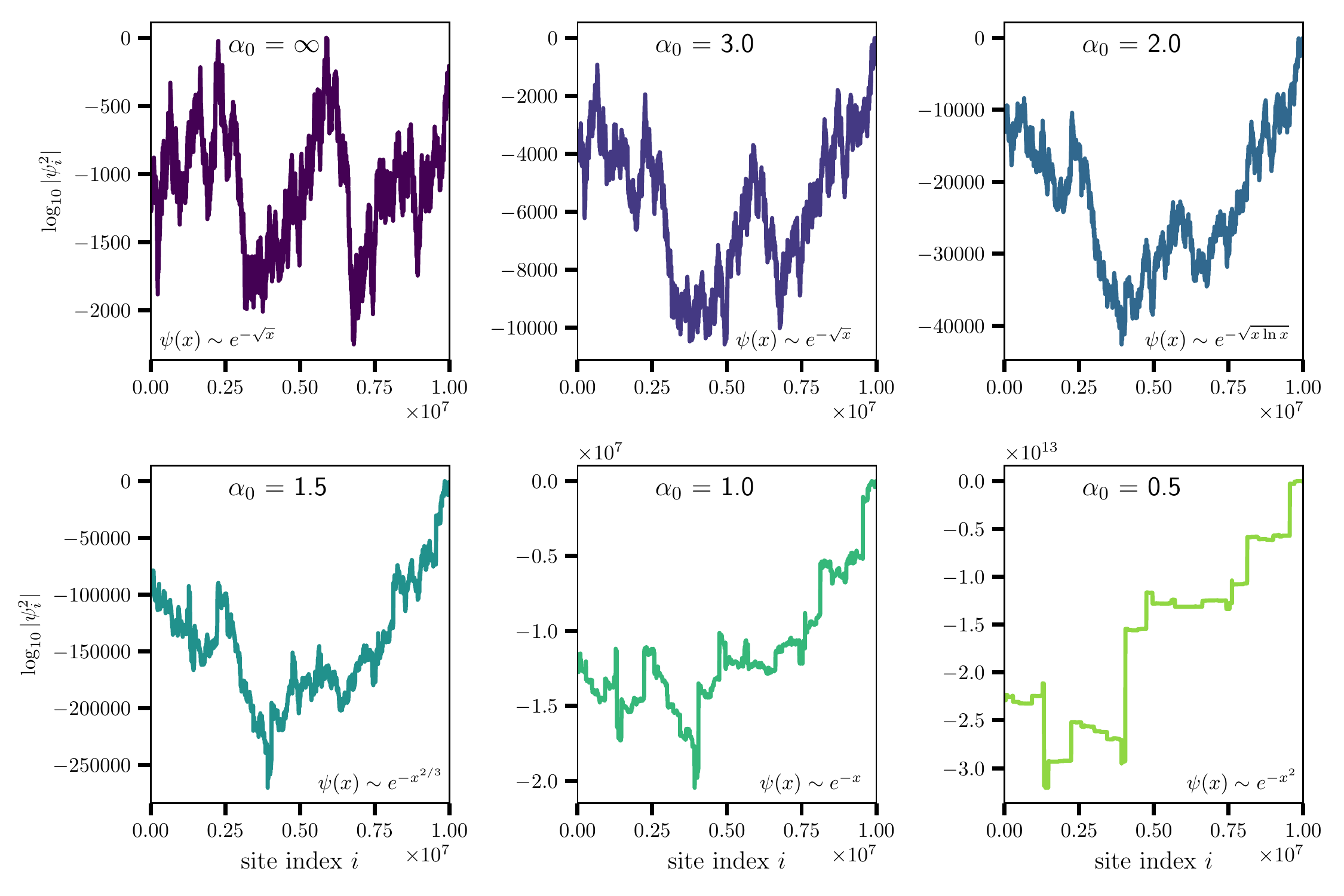}
	\caption{Examples of wave functions at energy $E=0$ on a system with $N = 10^7$ sites. The underlying realization of the $N-1$ random hoppings is the same for all sub-plots, i.e. their quantile functions are identical, before transforming them to the form in Eq.\ \eqref{eq:p_alpha}).
	We set $\alpha_\infty = 0$ in all cases above.
	The probability densities are chosen such that the hoppings $t$ have a median value $1$ for all choices of $\alpha_0$.
	Note the vertical axis is the log of wave function probability density, and its dynamic range increases rapidly as $\alpha$ is made smaller.
	%\akshay{Change lambda to alpha, remove IPR}
	}
	\label{FIG:1}
\end{figure}

The results in Eqs.\ \eqref{eq:CLT1} and \eqref{eq:CLT2} establish the nature of the wave function envelope -- it decays as a stretched exponential \begin{align}
\psi(r) \simeq \begin{cases} \exp  \left[ - \left(\frac{r}{r_0}\right)^{1/2}\right], \qquad \text{if } \alpha_0 > 2 \text{ and } \alpha_\infty > 2. \\
\exp  \left[ - \left(\frac{r}{r_0}\right)^{1/\alpha}\right] \qquad \text{otherwise.}
\end{cases}
\end{align}
The length scale $r_0$ is not constant -- it varies from one realization of disorder to the next, in accordance with the limiting Gaussian or alpha-stable distribution that it belongs to. 
%This is one example of the tunable manner in which the properties of the system my be adjusted by tweaking the parameters of the distribution of hoppings.
One case worth pointing out is $\alpha=1$, when the wave function decay is neither sub-exponential, nor super-exponential.
In this case, the distribution of $r_0$ is given by the positive part of $\mathcal{S}(1, 0, \pi, 0)$, which is better known in physics as the Cauchy-Lorentz distribution.
%For most other values of $\alpha \neq 1$, the L\'{e}vy alpha-stable distributions do not have simple closed-form expressions.

In Fig.\ \ref{FIG:1}, we show examples of wave functions at zero energy for a variety of values of the parameter $\alpha$.
It is apparent that for $\alpha > 2$, the wave function envelope decays less steeply than it does for $\alpha < 2$.
For $\alpha< 2$, the wave function envelope also has sharp discontinuities.

\section{Density of states and localization lengths} \label{sec:DoS}

In this section, we explore another effect of the varying the $\alpha_0, \alpha_\infty$ parameters in the hoppings -- that on the density of states $\rho(E)$ and integrated density of states $N(E)$, defined as \begin{align}
    N(E) \equiv \int\limits_{-E}^E \rho(E') \ \mathrm{d}E'.
\end{align}
Our technique here closely follows that of Eggarter and Riedinger  \citep{Eggarter1978}.
They map the quantum Hamiltonian to a classical random walk as follows.

Corresponding to the on-site wave function amplitudes $\psi_i$, the `self-energies' are defined as \begin{align}
\Delta_i \equiv t_{i-1} \frac{\psi_{i-1}}{\psi_i}.
\end{align}
In terms of these self-energies, the Schr\"{o}dinger equation, Eq.\ \eqref{eq:sch}, becomes \begin{align}
\Delta_{i+1} &= \frac{t_i^2}{E - \Delta_i}, \label{eq:SE1} \\
\Delta_{i+2} &= \left( \frac{t_{i+1}}{t_i} \right)^2 \Delta_i \left[ \frac{1-\frac{E}{\Delta_i}}{1 + \frac{E \Delta_i}{t_i^2}(1-\frac{E}{\Delta_i})} \right]. \label{eq:SE2}
\end{align}
The self-energies are useful because they are directly related to the integrated density of states.
Let $f_+$ be the fraction of positive self-energies in the sequence $\{ \Delta_0, \Delta_1, \cdots \Delta_i, \cdots \}$. It was shown by Schmidt in 1957  \cite{Schmidt1957} that \begin{align}
N(E) = 2f_+ -1. \label{eq:Ne}
\end{align}
When $E = 0$, it follows from Eq.\ \eqref{eq:SE1} that $\Delta_i \Delta_{i+1} < 0$, so the signs of the self-energies are alternately positive and negative.
Further, the term in the square brackets of Eq.\ \eqref{eq:SE2} may be ignored, and by taking the logarithm of both sides, Eq.\ \eqref{eq:SE2} may be interpreted as a discrete-time random walk \begin{align}
\ln \Delta_{i+2} = 2 \left( u_{i+1} - u_i \right) + \ln \Delta_i,
\end{align}
where $u_i \equiv \ln t_i$.
In the continuum limit, this is the Langevin equation \begin{align}
\frac{\mathrm{d}x}{\mathrm{d}t} = W_\alpha \zeta_\alpha(t), \label{eq:Langevin}
\end{align}
where the mapping $(\ln \Delta, i) \mapsto (x, t)$ converts a \emph{quantum} problem of self-energies on a discrete lattice to that of a \emph{classical} particle's position as a function of time.
The particle's $x(t)$ is governed by a unit-strength delta-correlated noise process $\zeta_\alpha(t)$.
The scale factor $W_\alpha$ is a dimension-full quantity that depends on the moments of the distribution of $u_i$.

When $\alpha > 2$, $\zeta_\alpha(t)$ has finite variance, and the stochastic process above is Brownian motion.
The scale factor is twice the variance of $u_i$, $W_\alpha = 2 \sigma_u$.
This directly corresponds to the case originally studied by Eggarter and Riedinger.
In this case, the Fokker-Planck equation corresponding to Eq.\ \eqref{eq:Langevin} is Fick's law of diffusion  \cite{Balakrishnan2008}, \begin{align}
    \frac{\partial}{\partial t} p(x, t)&= \frac{W_\alpha^2}{2} \frac{\partial^2}{\partial x^2} p(x, t), \label{eq:FPE}
\end{align}
where $p(x, t)$ is the probability density for the particle's position $x$ at time $t$.
The fundamental solution to this partial differential equation is a Gaussian kernel whose variance is proportional to time, leading to the well-known $\langle x^2 \rangle \propto t$ behavior.

But when $\alpha < 2$, the noise process $\zeta_\alpha(t)$ has infinite variance, and is referred to as `symmetric white alpha-stable noise'  \cite{Janicki1994}.
Such a stochastic process is known as a L\'{e}vy flight  \cite{Klages2008}, and unlike Brownian motion, can show discontinuous jumps.
It does not map to the Fokker-Planck equation (Eq.\ \eqref{eq:FPE}) above, and also possesses different scaling properties from that of a regular diffusive random walk.
In this situation, the scale factor depends on the parameters of the tails of the noise distribution, $W_\alpha = 2 \gamma(\alpha)$.
%In a L\'{e}vy flight, transport is super-diffusive, and the stochastic process Eq.\ \eqref{eq:Langevin} may be interpreted as the limiting case of the discrete random walk when the time increments $\Delta t \to 0$ \begin{align}
%\Delta x = x(t + \Delta t) - x(t) = W_\alpha (\Delta t)^{1/\alpha} \zeta_\alpha, \quad (\text{if } \alpha < 2)
%\end{align}
%where $\zeta_\alpha$ is a random variable drawn from the symmetric alpha-stable distribution $\mathcal{S}(\alpha, 0, 1, 0)$.

How does the random walk picture change when the energy is non-zero? It turns out it still holds, so long as $E$ is small. 
In the regime \begin{align}
E \ll \Delta_i \ll \tilde{t}^2/E, \label{eq:endpoints}
\end{align}
the terms in both the numerator and denominator of the square brackets of Eq.\ \eqref{eq:SE2} are small.
Here $\tilde{t}$ denotes a typical value of $t$.
The constraints enforced by Eq.\ \eqref{eq:endpoints} can be interpreted as boundary conditions for the random walk -- when $\Delta_i$ is large (comparable to $\tilde{t}^2/E$), the denominator in the square brackets ensures that $\Delta_{i+1}$ remains upper-bounded, and when $\Delta_i$ is smaller than $E$, then by Eq.\ \eqref{eq:SE1}, the alternating sequence of the sign of the $\Delta_i$'s is terminated by two consecutive positive values, with $\Delta_{i+1} \simeq \tilde{t}^2/E > 0$.
A new random walk then begins from this site.

The particle executes Brownian motion or a L\'{e}vy flight on a finite portion of the x-axis.
The domain is bounded on the right by an infinite potential barrier at $x_\text{max} = \ln (\tilde{t}^2/E)$, and on the left by an absorbing boundary at $x_\text{min} = \ln E$.
The particle commences its motion from a point just below $x_\text{max}$, and ends its motion when it reaches $x_\text{min}$.
Every iteration of this process corresponds to a single excess positive self energy in the alternating-sign sequence of $\{ \Delta_0, \Delta_1, \cdots \Delta_i, \cdots \}$.
The integrated density of states, which by Eq.\ \eqref{eq:Ne} is equal to the fraction of excess positive self-energies $(2 f_+ -1)$, is therefore inversely proportional to the mean time the particle spends moving freely in the interval $[x_\text{min}, x_\text{max}]$, i.e.\ \begin{align}
N(E) = \frac{1}{T_{\text{FP}}(\alpha, x_\text{max}, x_\text{min})}.
\end{align}

In the language of random walks, the quantity $T_{\text{FP}}(\alpha, x_\text{max}, x_\text{min})$ is known as the \emph{mean first passage time}.
It quantifies how long, on average, a particle under the influence of a stochastic process Eq.\ \eqref{eq:Langevin} and released at an initial position next to an infinite potential barrier $x_\text{max}$, survives before being absorbed at $x_\text{min}$.
The scale-invariance of this process implies that \begin{align}
T_{\text{FP}}(\alpha) &= \begin{dcases}
\tau_\alpha \left( \frac{x_\text{max} - x_\text{min}}{W_\alpha} \right)^2, \ & \alpha > 2 \enskip (\text{Brownian motion}), \\
\tau_\alpha \left( \frac{x_\text{max} - x_\text{min}}{W_\alpha} \right)^\alpha, \ & \alpha < 2 \enskip (\text{L\'{e}vy flight})
\end{dcases}
\end{align}
where $x_\text{max} - x_\text{min} = 2 \ln (\tilde{t}/E)$ is the length of the random walk, and $\tau_\alpha$ is the mean first passage time for the stochastic process Eq.\ \eqref{eq:Langevin} with unit strength over a unit interval, and is a number of order one.

\begin{figure}[t!]
	\centering
		\includegraphics[width=.5\textwidth]{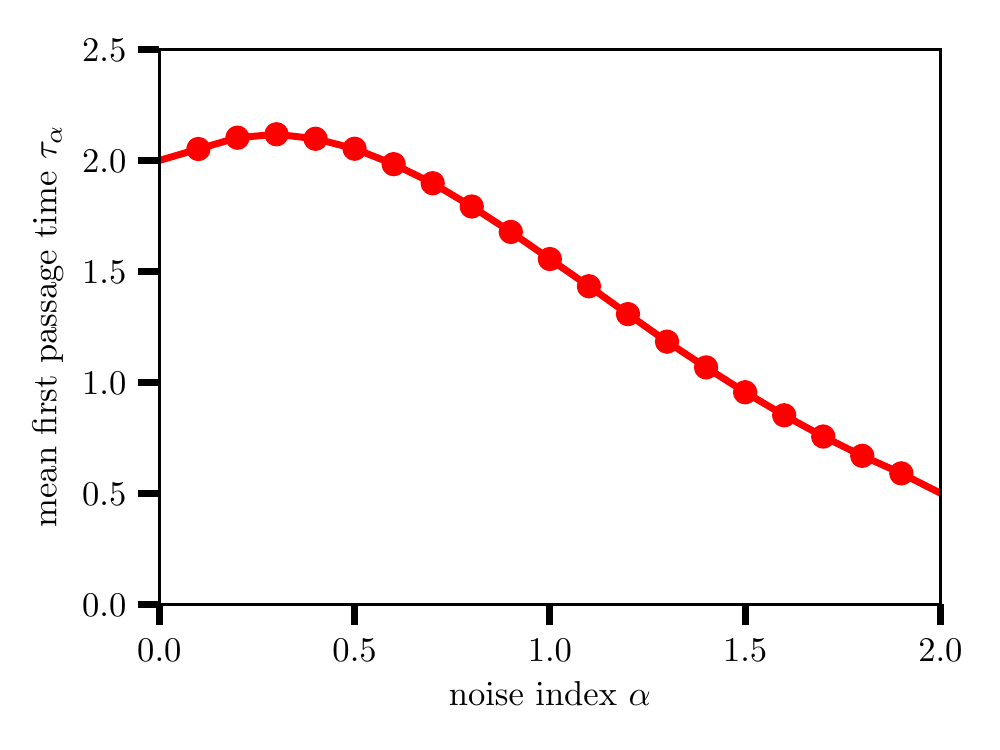}
	\caption{The mean first passage time $\tau_\alpha$ of the stochastic process in Eq.\ \eqref{eq:Langevin} with unit noise on a unit interval is calculated by numerical simulations. We discretize the Langevin equation with a time-step $\Delta t = 10^{-4}$ and average over $N=10^5$ trajectories for each value of $\alpha$. The error bars are smaller than the width of the points. For $\alpha > 2$ (not shown), $\tau_\alpha = 1$. In \ref{sec:App}, we discuss an alternative scheme leading towards obtaining this curve using analytic methods. This term is a dimensionless quantity that appears in the pre-factor for both the density of states $N(E)$ (Eq.\ \eqref{eq:DoS_th}) and localization length $\xi(E)$ (Eq.\ \eqref{eq:xiThou}).} %\akshay{Change lambda to alpha}}
	\label{FIG:2}
\end{figure}

For the Gaussian process in the case of Brownian motion ($\alpha > 2$), the Fokker-Planck equation in Eq.\ \eqref{eq:FPE} can be Laplace transformed in the time domain and integrated twice in space to obtain $\tau_\alpha = 1$ (see \ref{sec:App} for details).
However, for the L\'{e}vy flight ($\alpha < 2$), it is not so easy to obtain the first passage time analytically \cite{Koren2007, Dybiec2017, Metzler2019, Palyulin2019}.
%However, to the best of our knowledge there are no analytical results for $\tau_\alpha$: the case of a L\'{e}vy flight on finite 1-D domain with the particle being released adjacent to the reflecting boundary.
Nevertheless, it is fairly straightforward to compute $\tau_\alpha$ by numerical simulations  \citep{Krishna2020} (see Fig.\ \ref{FIG:2}).

We note that there appears to be a discontinuity in $\tau_\alpha$ at $\alpha = 2$, since above $\tau_\alpha$ appears to approach $1/2$ as $\alpha \to 2$ from below, but it is equal to a constant value of 1 for all values of $\alpha > 2$.
This discontinuity is of no major consequence; it is simply an artifact of the conventions used.
The Levy alpha stable distribution  $\mathcal{S}(\alpha=2, 0, \gamma=1, 0)$ corresponds to a Gaussian with variance 2.
For $\alpha>2$, we have considered a standard normal distribution with variance 1.

The desired result for the integrated density of states follows immediately as 
\begin{align}
N(E) = \begin{dcases}
\frac{\sigma_u^2}{|\ln E|^2}, \ & \alpha > 2, \\
\frac{[\gamma(\alpha)]^\alpha}{\tau_\alpha |\ln E|^\alpha}, \ & \alpha < 2.
\end{dcases} \label{eq:DoS_th}
\end{align}

The density of states $\rho(E)$ is intimately connected with another quantity of interest -- the localization length $\xi(E)$ via the Thouless relation \cite{Thouless1972} \begin{align}
\frac{1}{\xi(E)} &= \int\limits_{-\infty}^{\infty} \mathrm{d}E' \  \rho(E') \ln |E-E'| - \mu_u, \label{eq:Thou1}
\end{align}
where $\mu_u \equiv \langle u \rangle = \langle \ln t \rangle$ is the mean logarithm of the hopping terms. This quantity is divergent when $\alpha<1$.

For exponentially decaying wave functions with an envelope of the form $\psi(r) \sim \exp(-r/\xi)$, the localization length $\xi$ is a measure of how quickly the real-space probability amplitude attenuates.
The localization length is usually energy-dependent.
%The relationship above (Eq.\ \eqref{eq:Thou1}) is applicable whenever $\mu_u$ is finite, i.e.\ when ($\alpha > 1$).
As discussed in Sec.\ \ref{sec:zero}, when $\alpha > 1$, the state at zero energy is sub-exponentially localized, so $\xi(0) \to \infty$, and we obtain the sum rule \begin{align}
\int\limits_{0}^{\infty} \mathrm{d}E \  \rho(E) \ln(E) = \frac{\mu_u}{2} \label{eq:sumrule}.
\end{align}
Theodorou and Cohen \cite{Theodorou1976} showed that one can combine Eqs.\ \eqref{eq:Thou1} and \eqref{eq:sumrule}, to obtain the leading behavior of the localization length at non-zero energy.
 \begin{align}
\frac{1}{\xi(E)} &= 2 \int\limits_{0}^{E} \mathrm{d}E' \  \rho(E') \ln \left( \frac{E}{E'} \right) = \int\limits_0^E \mathrm{d}E' \  \frac{N(E')}{E'}, 
\end{align}
so that using Eq.\ \eqref{eq:DoS_th}, 
\begin{align}
\xi(E) &= \begin{dcases}
\frac{1}{\sigma_u^2} |\ln E|, &\alpha > 2 \\
\frac{(\alpha-1) \tau_\alpha}{[\gamma(\alpha)]^\alpha} |\ln E|^{\alpha-1}, &1 < \alpha < 2. 
\end{dcases} \label{eq:xiThou}
\end{align}

For $\alpha < 1$, the Thouless relation suggests that $\xi(E) = 0$ since $\mu_u = \infty$. 
We conclude that all the wave functions, across the entire spectrum, show a super-exponential decay.

\section{Conclusion}

This paper describes a non-interacting spinless fermionic system with random nearest neighbor hopping.
This model exhibits exact sub-lattice symmetry, even in the presence of disorder, and it is this exact symmetry that is largely responsible for its peculiar behavior.
The salient feature of our work is that the disordered hopping matrix elements $t$ are drawn from a probability density with a tunable sharp divergence of the form $1/t |\ln t|^{\alpha_0 + 1}$ as $t \to 0$, or a fat tail of the form $1/t |\ln t|^{\alpha_\infty + 1}$ at $t \to \infty$.
The parameter that controls the properties of the system is $\alpha \equiv \min(\alpha_0, \alpha_\infty)$
By extending an already known mapping between this quantum model and a classical discrete time random walk, we have been able to draw an exact parallel between the breakdown of universality and the violation of Fick's law of diffusion.

In the classical case, a fat-tailed noise process leads to super-diffusive behavior, with an anomalous exponent $\langle |x|^m \sim t^{m/\alpha} \rangle$, for moments $m< \alpha$.
In the disordered hopping model, a choice of hopping terms such that there is a large probability of either an exponentially small or a very large hopping term causes three major changes -- (i) the state at zero-energy decays as $\exp \left( - (r/r_0)^{1/\alpha} \right)$, which is super-exponential for $\alpha < 1$, (ii) the density of states shows a non-trivial logarithmic divergence at small energies, and (iii) the localization length also diverges non-trivially as the energy tends to zero.
Each of these three behaviors is tunable, depending on the divergence parameter $0 < \alpha < 2$ of the hopping distribution.
When $\alpha > 2$, we recover the previously known universal results that apply to all well-behaved distributions of hopping.
Table \ref{tab:table1} summarizes our systematic exploration of the entire parameter regime and encapsulates the central results of this paper.

\begin{table}[ht!]
\begin{minipage}{\textwidth}
\begin{tabularx}{0.88\linewidth}{@{} X | p{0.1mm} X X X @{}}
%\begin{tabular}{l | l l l l l l}
\toprule
\toprule
  & & $\bm{0 < \alpha < 1}$ &
  %\bigstrut & $\bm{\alpha = 1}$ & 
  $\bm{1 < \alpha < 2}$ & 
  %$\bm{\alpha = 2}$ & 
  $\bm{\alpha > 2}$ \\
\specialrule{.8pt}{2pt}{2pt}
%\multicolumn{7}{c}{statistics of hoppings $t$ and their distribution $p_\alpha(t)$} \bigstrut \\
%\midrule%\cmidrule(lr){1-6}
%  normalization $c_\alpha$ \bigstrut & & $\alpha(\alpha + 1)^\alpha$ & $2$ & $\alpha(\alpha + 1)^\alpha$ & $18$ & $\alpha(\alpha + 1)^\alpha$\\

%  median $t$ \bigstrut & & 1 & 1 & 1 & 1 & 1\\
%\specialrule{.8pt}{2pt}{2pt}
 
  \multicolumn{5}{c}{statistics of the logarithm of hoppings $u \equiv \ln t$}  \\
\midrule%(lr){1-6}
    
  mean $\mu_u$ &  & $\infty$ & finite & finite\\
 
  variance $\sigma_u^2$ & &
  $\infty$ & 
  $\infty$ & 
  %{$\begin{aligned}\frac{\alpha (\alpha+1)^2}{(\alpha-2)(\alpha-1)^2}\end{aligned}$} 
  finite
  \bigstrut \\ 
%\bottomrule
\specialrule{.8pt}{2pt}{2pt}
  \multicolumn{5}{c}{wave function of the band-center state ($E = 0$)}  \\ 
\midrule%(lr){1-6} 
  $\psi(r)$ & &
  {$\begin{aligned}e^{-(r/r_o)^{1/\alpha}}\end{aligned}$}& 
  %{$\begin{aligned}e^{-r/r_0}\end{aligned}$} & 
  {$\begin{aligned}e^{-(r/r_0)^{1/\alpha}}\end{aligned}$} & 
  %{$\begin{aligned}e^{-\sqrt{(r/r_0) \ln (r/r_0)}}\end{aligned}$}  \bigstrut & 
  {$\begin{aligned}e^{-\sqrt{r/r_0}}\end{aligned}$}\\
decay envelope  & & super-exponential & sub-exponential & sub-exponential\\

  distribution of $\frac{1}{r_0}$ %\footnote{in this row, $\mathcal{S}$ refers to the \emph{folded} symmetric alpha-stable distribution, and $\mathcal{N}$ to the folded Gaussian distribution, each with support only for positive values} 
  & &
  {$\begin{aligned}\mathcal{S}(\alpha, 0, \gamma(\alpha), 0)\end{aligned}$} 
  & 
  %{$\begin{aligned}\mathcal{S}(1, 0, \pi, 0)\end{aligned}$} & 
  {$\begin{aligned}\mathcal{S}(\alpha, 0, \gamma(\alpha), 0)\end{aligned}$} & 
  %{$\begin{aligned}\mathcal{N} \left( 0, \frac{c_\alpha}{2} \right) \end{aligned}$} \bigstrut & 
  {$\begin{aligned}\mathcal{N}(0, \sigma_u^2)\end{aligned}$} \\

\specialrule{.8pt}{2pt}{2pt}
  \multicolumn{5}{c}{integrated density of states $N(E)$ as $E \to 0$} \\
\midrule%(lr){1-6}
  $N(E)$ & & {$\begin{aligned}  \frac{b_\alpha}{|\ln E|^\alpha} \end{aligned}$} \bigstrut & 
  %{$\begin{aligned}\frac{b_\alpha}{|\ln E|}\end{aligned}$} & 
  {$\begin{aligned}\frac{b_\alpha}{|\ln E|^\alpha}\end{aligned}$} & 
  %{$\begin{aligned}\frac{b_\alpha \ln |\ln E| }{|\ln E|^2}\end{aligned}$} & 
  {$\begin{aligned}\frac{b_\alpha}{|\ln E|^2}\end{aligned}$} \\

  pre-factor $b_\alpha$  \bigstrut & &
  {$\begin{aligned}\frac{[\gamma(\alpha)]^\alpha}{\tau_\alpha}\end{aligned}$} %\footnote{see Fig.\ \ref{fig:tau} in Appendix \ref{apptau} for details on the mean first passage time $\tau_\alpha$} 
  & 
  %{$\begin{aligned}2\end{aligned}$} & 
  {$\begin{aligned}\frac{[\gamma(\alpha)]^\alpha}{\tau_\alpha}\end{aligned}$} \bigstrut & 
  %{$\begin{aligned}18\pm 2\end{aligned}$} & 
  {$\begin{aligned}\sigma_u^2\end{aligned}$} \\

%\bottomrule
\specialrule{.8pt}{2pt}{2pt}
  \multicolumn{5}{c}{spatial profile and localization length $\xi(E)$ at non-zero energy} \\
\midrule%(lr){1-6}
decay envelope  & & super-exponential & exponential & exponential\\

 $\xi(E)$ & &
 $0$ & 
 %$0$ & 
 {$\begin{aligned}\frac{(\alpha-1) \tau_\alpha}{[\gamma(\alpha)]^\alpha} |\ln E|^{\alpha-1}\end{aligned}$} \bigstrut &
 %{$\begin{aligned}\frac{1}{b_2} \frac{|\ln E|}{\ln |\ln E|}\end{aligned}$} & 
 {$\begin{aligned} \frac{|\ln E|}{\sigma_u^2} \end{aligned}$} \\
\bottomrule
\bottomrule
\end{tabularx}
\caption{Summary of key findings.
The hopping distribution $p(t)$ (Eq.\ \eqref{eq:p_alpha}) is parametrized by two exponents -- $\alpha_0$ and $\alpha_\infty$. However, the only quantity that matters is $\alpha \equiv \min(\alpha_0, \alpha_\infty)$.
For $\alpha > 2$, we recover the universal `Dyson' behavior, but for $\alpha < 2$, there is a variety of non-trivial physics.
The scale factor $\gamma(\alpha)$ referred to above is defined in Eq.\ \eqref{eq:gamma}, and the constant $\tau_\alpha$  is plotted in Fig.\ \ref{FIG:2}.
[A slightly more detailed version of this table is in Ref.\ \cite{Krishna2020}]
\label{tab:table1}
}
\end{minipage}
\end{table}

%The divergence is of the form $p(t) \sim 1/t \ln^{\alpha+1}(1/t)$.
While we have written our results in the language of an electronic model, the conclusions are directly applicable to the magnetic properties of spin-1/2 chains, as detailed in our earlier work  \citep{Krishna2020}.
The breakdown of universal low-temperature susceptibility directly follows from modified form of the density of states.
%While this work was being completed, we discovered that in his seminal work on random antiferromagnetic spin chains  \cite{Fisher1994}, Daniel Fisher did point out that if the couplings $J$ are distributed as $p(J) \sim 1/(J |\ln J|^x)$, then the properties of the system are dominated by weak links, and the behavior is expected to be non-universal. 
%However, there does not appear to have been a thorough investigation of the issue until now.
Our work also has strong connections with the strong-disorder renormalization group  \citep{Fisher1992, Fisher1994, McKenzie1996}.
While the infinite disorder fixed point was believed to be the only important one, Fisher \citep{Fisher1994} suspected that certain cases, where the couplings are dominated by weak links, would not behave universally.
Our investigation demonstrates the specific nature of this universality violation, and show the existence of an entire line of fixed points for $0 < \alpha < 2$.
One interesting feature is that this behavior can be triggered in two opposite kinds of situations -- either by having extremely weak or extremely strong nearest neighbor couplings.

In our earlier work \citep{Krishna2020}, we backed up our analytical predictions with two kinds of numerical analysis on systems of very large sizes ($N > 10^7$ sites).
Correlation functions obtained from transfer matrix calculations verified the functional form of the zero-energy state as well as the precise distribution of length scales $r_0$.
Strong-disorder renormalization group computations with ensemble averaging over hundreds of realizations corroborated our predictions for the density of states and the localization length to high precision, including the numerical pre-factors.
One peculiar feature of our model is the presence of extremely large variations in the hopping terms, over several thousands of orders of magnitude. 
This meant that the energies approached as small $10^{-10^8}$.
Our numerical calculations carefully implemented the arithmetic to avoid floating point errors over the full dynamic range of calculations.

From an experimental point of view, it might seem that such hoppings are challenging to realize in practice.
Nevertheless, it may be possible to construct such nearest neighbor couplings in synthetic lattices.
This may be achieved by distributing a chain of atoms randomly in a contrived potential of the appropriate  form.
The potential can be found by inverting the Schr\"{o}dinger equation, and is of the form $V(r)
 = V_0 + \left( c e^{2br} - e^{br} \right), \ \text{for } r > 0$, where $V_0$, $c$ and $b$ are positive constants \cite{Krishna2020}.
For negative $b$, this potential is known as the Morse potential  \cite{Morse1929}, used in studying the vibrations of diatomic molecules. 

One of the most remarkable and counter-intuitive aspects of disorder is universality. 
Disorder can ensue from a number of causes, such as positional randomness, crystallographic defects, impurity dopants  and so on.
However, the effect disorder has on quantum systems usually depends not on the origin or nature of disorder, but rather on the dimensionality and symmetries of the system.
It is this universality that makes simple theoretical models so powerful and applicable to real-world problems.
However, in this study, we have uncovered a non-trivial breakdown of universality that is driven purely by the choice of a probability distribution with fat tails.
This is similar to the origin of anomalous diffusion seen in L\'{e}vy flights and other kinds of non-Brownian stochastic processes.
%The non-trivial behavior of our model is entirely due to our choice of a probability distribution with fat tails.
%This is similar to the origin of non-universal anomalous diffusion seen in L\'{e}vy flights and other kinds of non-Brownian stochastic processes.
%While we have made a mapping between the low-energy properties of a quantum Hamiltonian and a first passage problem in classical non-equilibrium statistical mechanics, there are surely deeper connections waiting to be explored. 
Anomalous transport, L\'{e}vy flights, super-diffusion and allied topics are currently topics of great interest in the field of classical non-equilibrium statistical mechanics and mathematical physics \cite{Oliveira2019}, with applications to fields as diverse as biophysics \cite{Matthaus2009}, ecology \cite{Viswanathan2010} and finance \cite{Mariani2006}.
%Anomalous diffusion and L\'{e}vy flights are currently topics of active research, with ongoing efforts to push the boundaries of both mathematical theory and applications to other fields, including chemical kinetics, evolutionary biology, economics and finance.
In this context, it would be of interest to examine other quantum condensed matter systems where extreme value statistics and effects of large deviations cause a breakdown of universal behavior.
%Implications of this to systems in dimensions greater than one, or random graphs, \emph{maintaining the bipartite nature of the problem}, are also left for future work.

\section{Acknowledgments}

This work was supported by US Department of Energy, Division of Basic energy Sciences through Grant No. DE-SC0002140. RNB acknowledges the hospitality of the Aspen Center for Physics (supported by NSF), and of the Institute for Advanced Study, Princeton at different stages of the work.

After we submitted this manuscript, we learned that a similar approach had been used by Karevski et al \citep{Karevski2001} in the case of quantum magnets with broad disorder distributions. 
We thank the referee for pointing out this work, whose existence we were unaware of.

%\clearpage
%\begin{comment}
\appendix
\section{The mean first passage time} \label{sec:App}

It is known that under fairly general conditions, there is a correspondence between stochastic differential equations and Fokker-Planck equations.
This correspondence states that for any normalized stochastic differential equation with unit Gaussian noise $\zeta(t)$
\begin{align}
    \frac{\mathrm{d}x}{\mathrm{d}t} = f(x,t) + g(x, t) \zeta(t),
\end{align}
the probability density $p(x, t)$ evolves in accordance with the Fokker-Planck equation (FPE)
\begin{align}
    \frac{\partial}{\partial t} p(x, t)&=  -\frac{\partial}{\partial x} \left( f(x, t) p(x, t) \right) + \frac{1}{2 } \frac{\partial^2}{\partial x^2} \left( g^2(x, t) p(x, t) \right) \label{eq:appFPE}
\end{align}

For our situation in Sec.\ \ref{sec:DoS} (when $\alpha > 2$), the infinite potential barrier to the left of the origin can be modeled by a piecewise constant force field $f(x, t) = F \theta(-x)$, where the constant $F \to +\infty$, and $\theta(x)$ is the Heaviside step function. 
The noise is additive, so $g(x, t)$ can just be taken to be a constant. This simplifies the FPE considerably, \begin{align}
    \frac{\partial}{\partial t} p(x, t)&= -F \frac{\partial}{\partial x} \left( \theta(-x) p(x, t) \right)  + \frac{g^2}{2 } \frac{\partial^2}{\partial x^2} p(x, t). \label{eq:FPE2}
\end{align}

In our system, we have the initial condition $p(x, t=0) = \delta(x)$, the absorbing boundary at $x_0$ implies $p(x_0, t) = 0$.
The other spatial boundary condition is $p(x=-\infty, t) = 0$.

A quantity called the survival probability $p_s(t)$ can be defined as the integral of $p(x, t)$ over all space, and is the probability that the particle has not yet been absorbed at time $t$.
Common sense suggests that $p_s(t)$ monotonically decreases to zero from its initial value of $p_s(t = 0) = 1$.

The probability the particle first gets absorbed between time $t$ and $t + \Delta t$ is the difference in the survival probabilities at the two times, i.e. $p_s(t) -  p_s(t + \Delta t)$. The mean first passage time $\tau$ is the average of this quantity; thus \begin{align}
    \tau &= \lim\limits_{\Delta t \to 0} \sum\limits_t t \left( p_s(t) -  p_s(t + \Delta t) \right) = \int\limits_0^\infty \mathrm{d}t \ t \left( - \frac{d p_s(t)}{d t} \right) \\
    &= \int\limits_0^\infty \mathrm{d}t \ p_s(t) =  \int\limits_{-\infty}^{x_0} \mathrm{d}x \int\limits_0^\infty \mathrm{d}t \ p(x, t) = \int\limits_{-\infty}^{x_0} \mathrm{d}x \ \tilde{p}(x, s=0),
\end{align}
where $\tilde{p}(x, s)\equiv \int\limits_0^\infty \mathrm{d}t \ e^{-st} p(x, t)$ is the unilateral Laplace transform.
Calling $q(x) \equiv \tilde{p}(x, s=0)$ for convenience, so that the mean first passage time is just a definite integral of a single-variable function, \begin{align}
    \tau &= \int\limits_{-\infty}^{x_0} \mathrm{d}x \ q(x) \label{eq:tauint}
\end{align}

Taking the unilateral Laplace transform in time of Eq.\ \eqref{eq:FPE2}, and setting $s=0$, we end up reducing a partial differential equation to an ordinary differential equation \begin{align}
    - \delta(x) &=
    -F \frac{d}{d x} \left( \theta(-x) q(x) \right)  + \frac{g^2}{2 } \frac{d^2}{d x^2} q(x), \quad \text{which when integrated once,} \label{eq:FPE3} \\
    -\theta(x) &= -F \theta(-x) q(x) + \frac{g^2}{2} \frac{d}{d x} q(x) 
\end{align}

The solution to this ODE (with boundary conditions $q(-\infty) = q(x_0) = 0$) is \begin{align}
    q(x) = \begin{dcases}
    \frac{2 x_0}{g^2} \exp \left( \frac{2F x}{g^2} \right), & -\infty < x < 0 \\
    \frac{2 (x_0 - x)}{g^2}, &0 < x < x_0.
    \end{dcases}
\end{align}
And it follows by doing the integral in \eqref{eq:tauint}, \begin{align}
    \tau = \frac{x_0}{F} + \frac{x_0^2}{g^2}.
\end{align}

For the random walk in Sec.\ \ref{sec:DoS}, we set $F \to \infty$, and $x_0$ and $g$ are normalized to 1, so $\tau_\alpha = 1 \ \forall \ \alpha > 2$.

%\akshay{Referee comments: "For alpha = 2, is tau = 0.5 as suggested by the solid line?
%Maybe provide a comment on the discontinuity of tau, if any?"}

%\akshay{Response: For $\alpha > 2$, $\tau_\alpha = 1$. This discontinuity is because the Levy alpha stable distribution  $\mathcal{S}(\alpha=2, 0, \gamma=1, 0)$ corresponds to a Gaussian with variance 2. This is not a unit normal distribution.}

When $0 < \alpha < 2$, the situation is different.
For a symmetric L\'{e}vy noise process, Eq.\ \eqref{eq:appFPE} is replaced by the generalized Fokker-Planck equation \cite{Denisov2009},
\begin{align}
    \frac{\partial}{\partial t} p(x, t)&=  -\frac{\partial}{\partial x} \left( f(x, t) p(x, t) \right) + \frac{\partial^\alpha}{\partial |x|^\alpha} \left( g^\alpha(x, t) p(x, t) \right), \label{eq:fracFPE1} %\\
    %\frac{\partial}{\partial t} p(x, t)&=  -\frac{\partial}{\partial x} \left( f(x, t) p(x, t) \right) - \frac{D^\alpha_+ + D^\alpha_-}{2 \cos (\pi \alpha/2)} \left( g^\alpha(x, t) p(x, t) \right) \label{eq:fracFPE2}
\end{align}
implying that the differential equation \eqref{eq:FPE3} to be solved becomes \begin{align}
    - \delta(x) &=
    -F \frac{d}{d x} \left( \theta(-x) q(x) \right)  + g^\alpha \frac{d^\alpha}{d |x|^\alpha} q(x). \label{eq:fracFPE4}
\end{align}

In Eq.\ \eqref{eq:fracFPE1}, $\frac{d^\alpha}{d |x|^\alpha}$ denotes the Riesz fractional derivative \cite{Herrmann2014}, which is defined in terms of the Fourier transform as \begin{align}
    \frac{d^\alpha}{d |x|^\alpha} f(x) \equiv \int\limits_{-\infty}^\infty \frac{\mathrm{d}k} {2\pi} |k|^\alpha e^{-ikx} \int\limits_{-\infty}^\infty \mathrm{d}x' f(x') e^{ikx'}.
\end{align}
%In Eq.\ \eqref{eq:fracFPE2}, an alternative expression for the fractional derivative is used, which is the Riemann-Liouville operators $(D^\alpha_\pm)$.

The solution to Eq.\ \eqref{eq:fracFPE4} should enable one to find the mean first passage time $\tau_\alpha$ for the L\'{e}vy noise case.
This provides the foundation for an alternative method to obtain the curve in Fig.\ \ref{FIG:2}.
However, we are unable to proceed and obtain a closed-form expression for $\tau_\alpha$.

\clearpage
%\end{comment}

\bibliography{cas-refs}

\end{document}